\documentstyle[psfig]{laa}
\psfull
\def\ltsima{$\; \buildrel < \over \sim \;$}
\def\lsim{\lower.5ex\hbox{\ltsima}}
\def\gtsima{$\; \buildrel > \over \sim \;$}
\def\gsim{\lower.5ex\hbox{\gtsima}}
\def\cmdue{~cm$^{-2}$}
\def\deg {$^\circ$}

\title{Search and analysis of small scale structures in two X--ray clusters
of galaxies}
\author{D.~Lazzati\inst{1,2} 
\and S.~Campana\inst{1}
\and P.~Rosati\inst{3}
\and G.~Chincarini\inst{1,2}
\and R.~Giacconi\inst{3,2}}

\begin{document}

\offprints{D.~Lazzati}

\institute{
{Osservatorio Astronomico di Brera, Via E. Bianchi 46, I-23807 Merate,
Italy}
\and{Universit\`a degli studi di Milano, Via Celoria 16, I-20133 Milano,
Italy}
\and{European Southern Observatory, Karl Schwartzschild Strasse, Garching,
Germany}
}
\date{Received ; Accepted}

\maketitle

\begin{abstract}

We present a refinement of the wavelet analysis technique for the
detection and characterisation of small scale features embedded in a
strongly varying background. This technique handles with particular care
the side effects of non-orthogonality in the wavelet space which can
cause spurious detections and lead to a biased estimate of source parameters.
This novel technique is applied to two ROSAT PSPC pointed observations of
nearby clusters of galaxies, A1367 and A194. We find evidence that
the case of A1367 is not unique and that galaxy-scale X--ray emission
could be a quite common property of clusters of galaxies. We
detect 28 sources in the field of A1367 and 26 in the field of A194.
Since these numbers are significantly larger than those expected from
the $\log N - \log S$ relation in the field, most of the sources are
expected to be associated with the cluster itself and indeed several
identifications with galaxies are possible.
In addition, CCD observations have revealed that two X--ray sources in the
field of A194, classified as extended by the multi-scale analysis, are
very likely associated with two background galaxy clusters at intermediate
redshift.


\end{abstract}
              
\section{Introduction}
\label{intro}

The study of the X--ray emission from clusters of galaxies plays a central 
role in understanding the origin of the intra-cluster medium (ICM), 
its interaction with the cluster galaxies and the physics related to the merging
processes in the ICM.
A first detection of X--ray emission from the single galaxies in a
cluster, beyond Virgo, was made in A1367 by Bechtold et al. (1983,
hereafter BAL) using the High Resolution Imager (HRI) and Imaging 
Proportional Counter (IPC) instruments on board {\it Einstein}.
Following this work, optical observations have shown that in such
X--ray emitting galaxies high temperature coronal gas coexists with
cold HI gas, although the latter is found to be significantly
underabundant when compared to X--ray quiet cluster galaxies (Chincarini
et al. 1983).

The detection of individual galaxies in clusters has been limited so
far by the angular resolution of the X--ray instruments.  Although the
next generation of X--ray telescopes (e.g. AXAF, JET--X, XMM) will
facilitate these studies, the use of a flexible detection algorithm is
the main ingredient to generate reliable catalogs of X--ray emitting
galaxies in clusters, by separating the small scale features from the
strong diffuse cluster emission, and provide accurate measurements of
characteristic sizes and fluxes.

A previous attempt to apply a wavelet transform-based algorithm
to a new pointed ROSAT observation of A1367 was carried out by 
Grebenev et al. (1995, hereafter GAL).
They generated a catalog of point sources and discuss possible identifications
of optical counterparts in the cluster field. 
In this work we extend the GAL analysis and present an improvement of
the wavelet detection technique which is designed to minimise the side
effects of the transform and to deal with the a strongly-varying
background and source confusion.
The method is then applied to the search for small-scale X--ray emission
in two clusters of galaxies, A1367 and A194.

Our work is organised as follows: in section \ref{metodo} we briefly 
outline the wavelet algorithm while in section \ref{analysis} we detail 
our analysis. Results of individual clusters are given in section 
\ref{individual}; section \ref{concl} contains a brief summary.

\section{The Method}
\label{metodo}

The method we present is based on the wavelet transform, a mathematical
tool which has been recently used for multi-scale data analysis in
different astronomical applications.
For the complete theory of wavelet 
operators, we  refer to some of the extensive reviews published 
in the literature, e.g. GAL and Slezak et al. (1995). The wavelet
transform (hereafter WT) of a real function $f$ is defined as the
convolution product (more precisely as the scalar product) 
between the function and a class of
analysing wavelets $\{ \psi_{a,b} \}$, all derived by translation and
dilation from a single function, generally complex (but in our case
real), called mother wavelet, which has to fulfill the condition
(in the discrete approximation):

\begin{figure*}[!htb]
\centerline{\psfig{figure=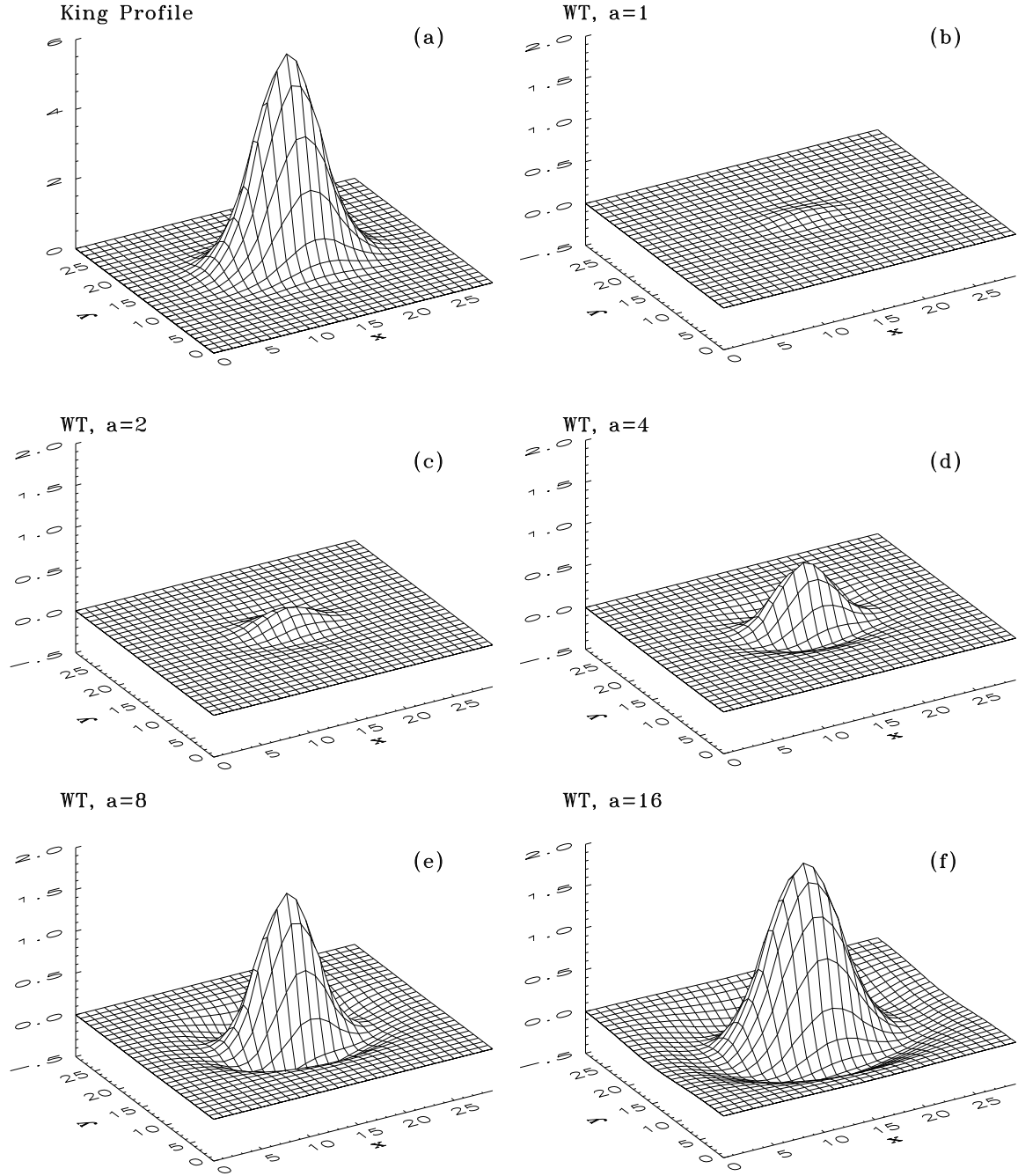,width=15cm}}
\caption{Example of non-orthogonality in scale: the six panels show
an elliptical King profile [panel (a)] and its WT at five increasing scales
[panels (b)-(f)]. It can be seen that the coefficients of the small scale
transforms [panel (b)-(d)] are not equal to zero, as discussed in the text.
The vertical axes of panels (b)-(f) have the same dimensions.}
\label{kingps}
\end{figure*}

\begin{equation}
\sum_{i,j} \psi_{i,j} = 0  \label{cond}
\end{equation}
\noindent for a two-dimensional transform.

The discrete approximation of the WT can hence be written as a convolution 
in $l^2$:
\begin{equation}
w_{i,j,a_n}={1\over {a_n^2}} \sum_{k,l} f_{k,l} \; \psi_{{{i-k}\over{a_n}},
{{j-l}\over{a_n}}}  \label{wtdef}
\end{equation}
\noindent
where, for a proper sampling, the scales $a_n$ must be chosen according to:
\begin{equation}
a_n=(2^m)^n \qquad\qquad\qquad m \in {\bf Z} \label{sampl}
\end{equation}
\noindent
with $m$ fixed.

As pointed out by many authors, the most appealing property of the WT
resides in its capability of decomposing $f$ in a certain number of
functions, each representing the features of $f$ at a given scale,
leaving unaffected the positional information. 

The WT has another great advantage, it gives an exact estimate of the
significance of correlated multi-pixel enhancements, regardless of the amplitude
of the enhancement, even in the presence of non-flat background
components.  Moreover, this process does not require an estimate of the
local background since the variance in the transform pixel by pixel can
be calculated without assuming a background model.

The formula for the background variance is given in GAL:
\begin{equation}
\sigma^2_{i,j,a_n}={1\over{a_n^4}}\sum_{k,l} {\sigma_f^2}_{k,l} \;
\psi^2_{{{i-k}\over {a_n}},{{j-l}\over {a_n}}}     \label{sigma}
\end{equation}
\noindent
where we have used the symbol $\sigma$, typical of the Gaussian
distribution, because it can be shown that the single WT coefficients are
distributed according to a Gaussian statistics with standard deviation
given by Equation \ref{sigma} (Lazzati 1996). This result is exact if
$f_{i,j}$ have a Gaussian distribution and is a very good approximation for a
Poisson distribution if the mean number of counts is sufficiently large
 ($\gsim 20$) or
the scale $a_n$ is sufficiently large. 

Despite all these properties, the WT has a major disadvantage which
usually tends to be neglected (e.g. GAL) or overemphasised (e.g. Pando
\& Fang 1996): the wavelet bases commonly adopted for astronomical purposes
are {\sl not orthogonal, neither for rescaling nor under translations}.

\begin{figure}[!hbt]
\centerline{\psfig{figure=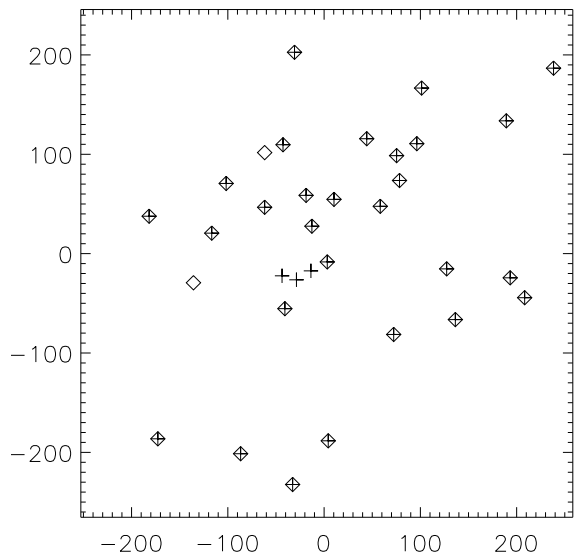,width=9.cm} {\hfil}}
\caption{Sources detected in the field of A1367: crosses and diamonds
represent detections before and after the King profile
subtraction, respectively. The center of the cluster is at (0,0). 
See the text for the relevance of crosses without diamonds
and diamonds without crosses.} 
\label{conf}
\end{figure}

\subsection{Non-orthogonality in space}
\label{correlation}

In the analysis of astronomical images, the non-orthogonality 
under translations has the effect
of producing a correlation between the uncertainties on WT-space coefficients.
In the mere identification of sources this is not a major problem, since the
statistical significance of detections comes from the value of a single 
coefficient. However, it is mandatory to take into account this effect
if standard fitting procedures have to be used.

Problems arise for near sources since the tail of the first
source can affect the significance of the second. This is not an easy
problem, but it has been shown (Rosati 1995; Rosati et al. 1995) that WT
based detection algorithms deal the confusion problem better than other
techniques commonly used in X--ray astronomy, such as the sliding box
technique.

Given the fact that the number of independent points in a single-scale
transform is not known, it is not possible to derive analytically 
the expected number of spurious sources
in the detected sample, despite the Gaussian character of the statistic.
Simulations must be performed to assess the significance of the detection as
in GAL. These simulations however are rather difficult, due to the lack 
of a consistent model for the background component (we call background
also the large-scale emission from the ICM).
In this work we prefer to
give the significance of each single source, i.e. we quote for each
source the probability (in units of $\sigma$) of spurious detection.

\subsection{Non-orthogonality in scale}
\label{orthscale}

A more serious problem is the non-orthogonality among different scales
which causes large scale features to affect the coefficients of the
small scale transforms. This effect is particularly important when we
try to detect small and faint scale features embedded in a strong,
large scale component.

As an example, we can consider the WT of a King profile modeling the
cluster diffuse X--ray emission. It can be noticed that the transform
has a maximum  at its natural scale, but all the other scales have
non-zero coefficients (see Figure~\ref{kingps}).  Since the operator is
linear (see Equation \ref{wtdef}), the transform of the superposition
of a small scale feature (e.g. a galaxy) on the King profile in
Figure~\ref{kingps} will be the sum of the transforms of the King
profile and that of the small feature. The way how the significance of
a small-scale feature is altered by the presence of the King profile
depends on the position, since the transform of the profile is positive
in the center and negative in an annulus around it.  As a result, we
overestimate the significance of small scale structures near the center
of the cluster and underestimate that of sources in the outskirts.

\section{The Analysis}
\label{analysis}

We consider here the ROSAT  Position Sensitive Proportional Counter (PSPC) 
observations of two clusters of galaxies (A194 and A1367), 
characterised by relatively small distances and long exposure times.
In Table~\ref{sampletab} we report the properties of these clusters 
and of their ROSAT observations.

\begin{table}[!hbt]
\caption{Properties of clusters and of ROSAT observations.} 
\label{sampletab}
\centerline{
\begin{tabular}{l|ccccc}
Name    & $R^a$ & Redshift      & $BM\,^b$      & Obs. ID       & Exp. (s) \\ 
\hline
A194    &   0   & 0.0178        & II            & rp800316      & 16317    \\
A1367   &   2   & 0.0215        & II-III        & rp800153      & 18224    \\
\end{tabular}}

\footnotesize{
$^a$ Richness class

$^b$ Bautz--Morgan type
}
\end{table}

\subsection{Data reduction}
\label{dr}

The reduction of the ROSAT PSPC data has been performed with the method 
described by Snowden et al. (1992; 1993; 1994) and Plucinsky et al. (1993).
The complete treatment 
takes into account the long and short term enhancements of background,
the solar afterpulse and particle contamination, and the energy-dependent
correction for vignetting and exposure time.

In order to oversample the PSPC Point Spread Function (PSF), the standard 
ESAS software (Snowden 1995) has been generalised to produce images with 
smaller pixel size (Lazzati 1996). 
In the following we analyse images in the hard band ($0.4-2.0$~keV)
with a pixel size of 8 arcsec.

\begin{figure}[!hb]
\centerline{\psfig{figure=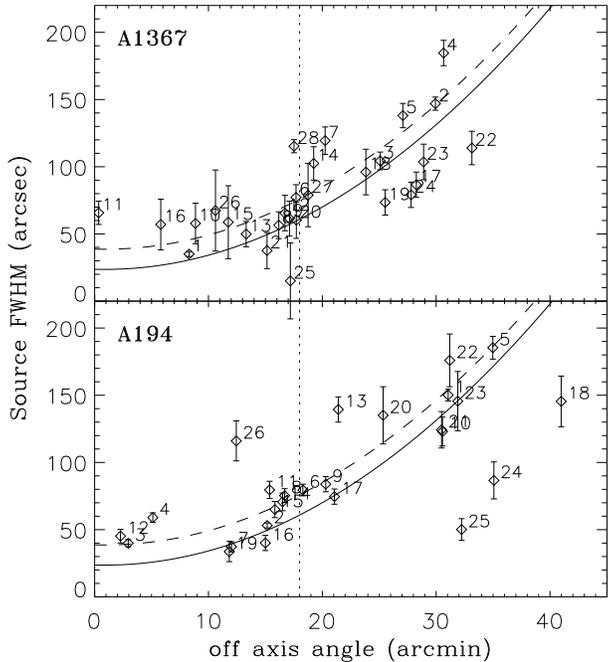,width=8.cm} {\hfil}}
\caption{FWHM of detected sources compared to the PSPC PSF FWHM  
depending on the off-axis angle. The model for the PSF (solid line) 
is accurate up to $\theta \leq 18'$. Dashed lines represent $3\,\sigma$ 
confidence level for the PSF model. A1367 field is the upper panel, 
A194 the lower.}
\label{psfsou}
\end{figure}

Our source detection technique utilises the discrete WT based on the
multi-resolution theory (Farge 1992) and the ``\`a trous" algorithm
(Bijaoui \& Giudicelli 1992). The latter uses the `mother' developed by
Slezak et al. (1995), which can be analytically approximated with 
the difference of Gaussians:

\begin{eqnarray}
\psi_{j,a,b} & = & 2\pi \Big[ A\,\sigma_A^2\, G(x,y,a,b,2^j\sigma_A) - \nonumber \\
& - & B\,\sigma_B^2\,G(x,y,a,b,2^j\sigma_B)\Big]
\label{mothereq}
\end{eqnarray}

\noindent where

\begin{equation}
G(x,y,x',y',\sigma) = \frac{1}{2\pi \, \sigma^2}\,e^{\frac{(x-x')^2+(y-y')^2}{2\sigma^2}}
\label{defgau}
\end{equation}

\noindent $A$ and $B$ are the normalisation factors and $\sigma_a$ e $\sigma_b$
the widths of the two Gaussians.

\begin{figure}[!ht]
\psfig{figure=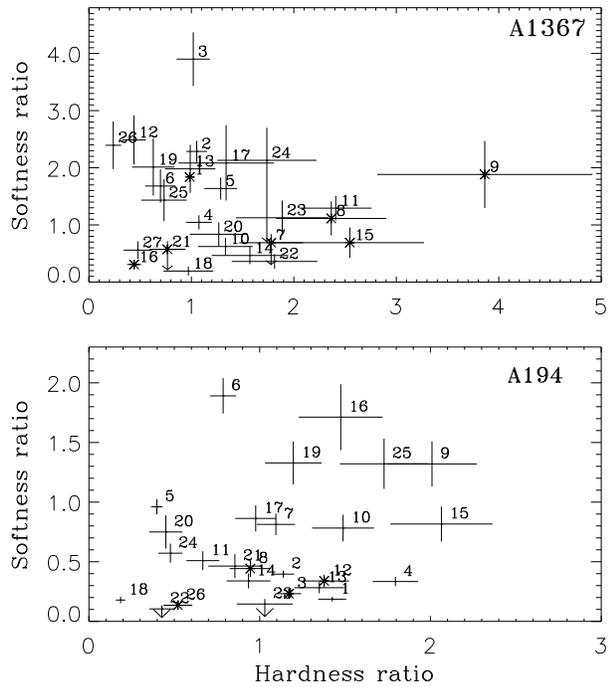,width=8.0cm}
\caption{Color--color diagrams for the most luminous sources in the
field of A1367 (upper panel) and A194 (lower panel).
The softness ratio is defined as the ratio between the flux
in the bands $(0.11-0.28)$~keV and $(0.28-1.2)$~keV, while the hardness 
ratio as the ratio between bands $(1.2-2.04)$~keV and $(0.28-1.2)$~keV.
We expect sources associated with cluster objects to be hard due to
the low energy absorption of the galactic interstellar medium ($N_H\sim 
5\times10^{20}$\cmdue). Sources associated with cluster galaxies 
are marked with an asterisk.}
\label{coldiag}
\end{figure}

\subsection{Large scale emission}
\label{lse}

In order to correct for effects of the WT non-orthogonality (see 
section~\ref{orthscale}) and remove the cross-talk between the largest
dominant scale and the small scale features, we fit and subtract an elliptical 
two-dimensional King profile from the original data.
The fitting procedure is performed after masking regions flagged as 
compact sources or substructures by a first run of the detection algorithm.
The best fit model is then subtracted and the residual image 
used for further analysis.

A great amount of study has been devoted to test the reliability
of the two-dimensional fitting. Two possible solutions were explored 
in order to overcome the poor statistics in the external regions of the 
cluster. A first simple
approach is to fit the image assuming Poissonian errors. 
This has the disadvantage of introducing a bias in the fitting procedure
since negative deviations are weighted more than positive ones, thus
producing an annulus with an excess of counts in the residual image.

\begin{figure*}[!ht]
\psfig{figure=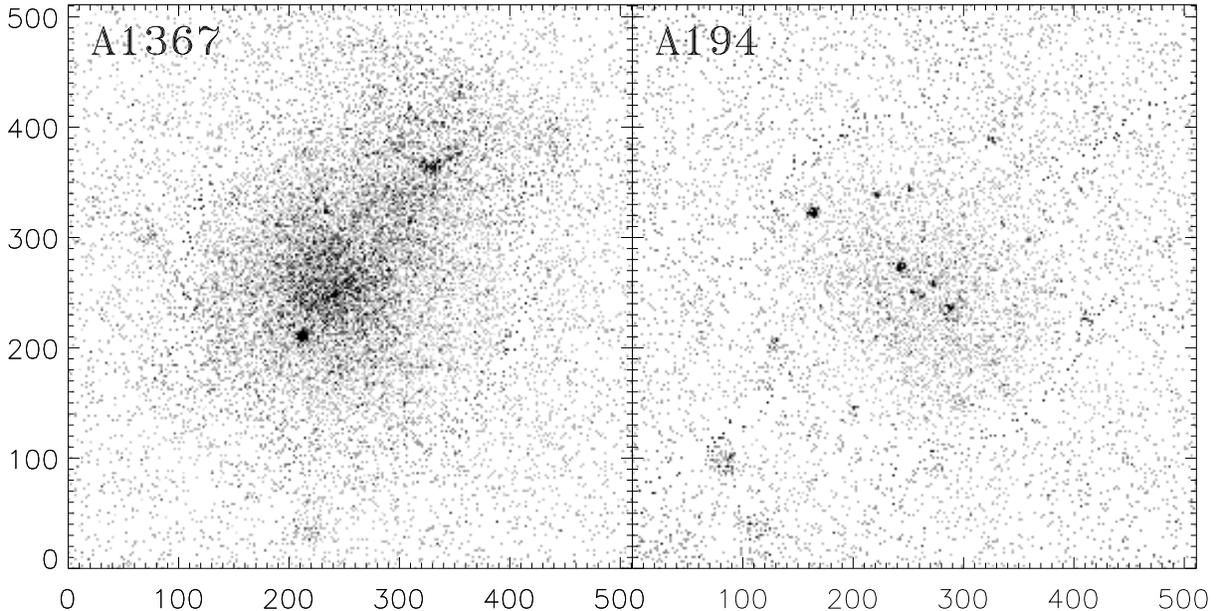}
\caption{X--ray images of the two clusters analysed in the text. The
fields are $68'$ across, the pixel size is $8''$.}
\label{xps}
\end{figure*}

In order to obtain a better statistics without compromising the 
resolution, we have preferred to fit the King model to the original image
smoothed with a bidimensional Gaussian (see also BAL).
The reliability of this non standard procedure has been confirmed
through a large set of simulations from which we have verified that the
fitted and the true parameters are in good agreement as long as the
size of the smoothing function is comparable to the detector PSF. 

The outlined procedure, when compared to a cluster profile subtraction and 
detection in real space, has the following advantages:\\
a) the WT space is less affected by 
   residuals from the large scale subtraction;\\
b) the WT algorithm is efficient also in cases where the standard deviation
   is not constant over the analysed field;\\
c) the WT algorithm proves to be very powerful for multi-scale source detection.

The importance of the cluster profile subtraction, i.e. the fact that
we have to properly take into account the background signal to assess
correctly the significance of detections, is illustrated in
Figure~\ref{conf}. We show the sources detected at a $4\,\sigma$ level
in one of the most studied clusters of galaxies, A1367 (BAL; GAL).
Crosses and diamonds represent the sources detected before and after the 
subtraction of the fitted King profile, respectively.
We note that the majority of the sources are detected in both cases, 
however in the central region (where the cluster emission is higher) 
differences can be found. The three central detections marked with a simple
cross have a significance lower then $4\,\sigma$ but were enhanced above
this threshold because of the coincidence 
with the peak of the cluster emission.
The two sources marked with open diamonds are significant ($>4\,\sigma$) 
features suppressed by the negative wings of the cluster emission transform
(see Figure~\ref{kingps}). This example elucidates the importance of taking 
into account the
diffuse cluster emission which can seriously affect the significance of
source detections in the central regions.

\begin{figure}[!t]
\psfig{figure=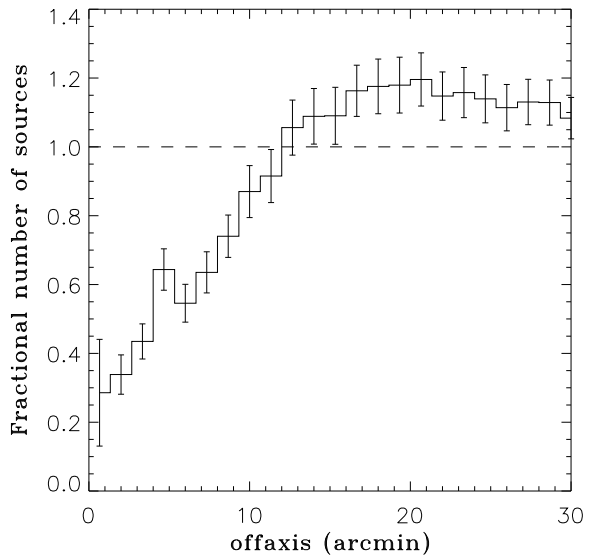,width=8cm}
\caption{{Ratio between the surface density of detected sources 
with and without the King profile subtraction described in the text.
Data are obtained from a set of 100 simulations of an A1367-like image 
with 150 compact sources superimposed.} \label{refps}}
\end{figure}

\begin{figure*}[!ht]
\centerline{\psfig{figure=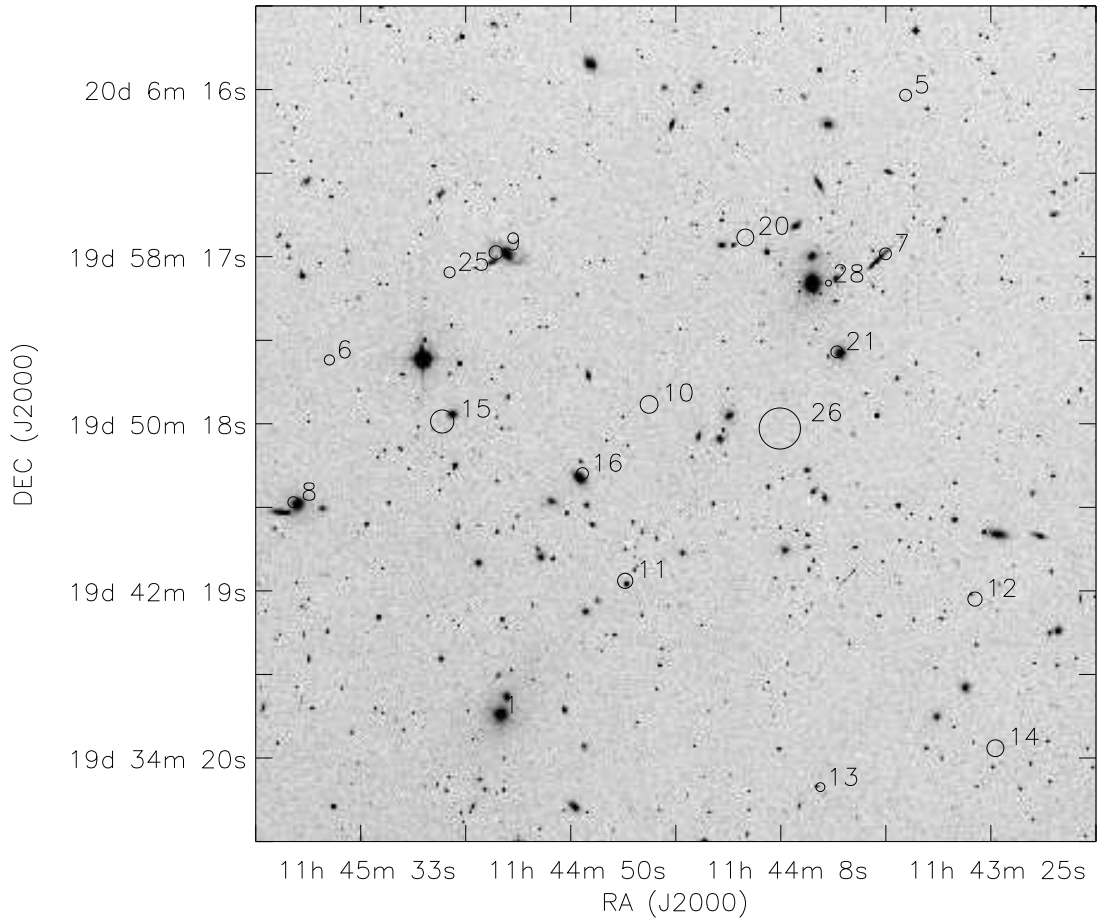,width=15cm}{\hfil} }
\caption{X--ray sources detected in the central $40'\times40'$ of A1367
field overlaid on the relevant Digital Sky Survey (hereafter DSS) 
plate. The radii of circles represent 
the $3\,\sigma$ confidence radius on the central position of the X--ray source.
The image has been corrected for boresight errors ($\sim 14''$; see text).}
\label{a1367poss}
\end{figure*}

To test the influence of this effect, we have simulated a set of 100 
images of an A1367-like cluster with the superposition of 150 compact sources 
distributed along a power-law $\log N - \log S$ with index $\alpha=-2.5$. 
We have performed 
the source detection with and without the King profile subtraction.
In Figure~\ref{refps}  the ratio of the radial surface density
profiles of detected sources in the two realisations is shown.
in the central region of the image the number of sources detected without
the King profile subtraction is up to three times greater than that 
of sources detected with our method. In the outskirt of the cluster, without
a careful handling of the negative wings, up to $20\%$
of the sources are missed.

This problem has been faced in a recent paper by Damiani et al. (1997a, b)
who calculate background maps at each scale by smoothing the image with different
widths of the window. With this method, large features are seen as background 
at the lower scales, whereas small scale features are averaged in the upper scales.
However the influence of the cross talks between different scales cannot
be solved.

In a completely different approach,
Bijaoui et al. (1995), Rue \& Bijaoui (1996), 
Pislar et al. (1997) have discussed 
a restoration algorithm based on WT coefficients 
which does not need any parametrisation of the source shape.
Similarly, Vikhlinin et al. (1997) subtract the small scale features 
from the WT before the computation of the larger scale
coefficients.

\subsection{Characterisation of detected sources}
\label{char}

To characterise the detected sources, we developed a multi-scale fitting 
method (Lazzati et al. 1997; Campana et al. 1997). 
The basic formula of this method is the WT of a Gaussian source that, following
equation~\ref{defgau} and given the mother of equation~\ref{mothereq} 
can be written:

\begin{eqnarray}
\tilde f(a,b) & = & 2\pi\, I_0 \, \bigg[ A\,\sigma_A^2 \,G \Big(a,b,x',y',\sqrt{\sigma_2+2^{2j}\sigma_A^2}\Big) - \nonumber \\
& - & B\,\sigma_B^2\,G\Big(a,b,x',y',\sqrt{\sigma^2+2^{2j}\sigma_B^2}\Big)\bigg]
\end{eqnarray} 

\noindent where $I_0$ is the total number of source 
counts and $\sigma$ its width.

In this approach the WT of a Gaussian source is fit 
to the image transform at all scales simultaneously.
This allows us to characterise in a single operation 
the multi-scale and two-dimensional behaviour of the detected sources, 
using all the available information. Thus we obtain the most accurate
determination of the interesting parameters: position, size and count-rate.
Unfortunately, due to the high spatial correlation of WT parameters, 
especially at the largest scales, a simultaneous fit of all coefficients would
be time-expensive and redundant. 
Moreover, uncertainties of adjacent pixels are correlated and standard 
fitting techniques should not be used.

\begin{figure*}[!ht]
\centerline{\psfig{figure=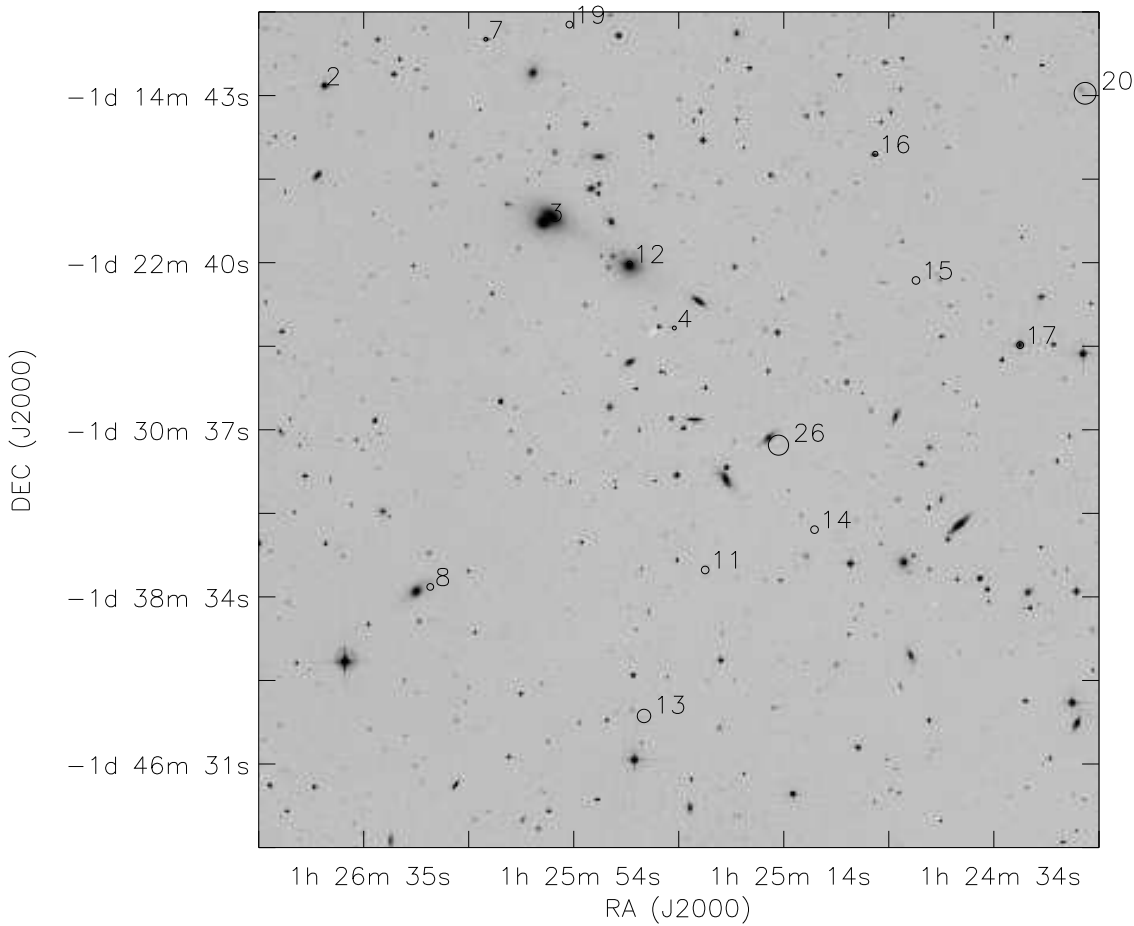,width=15cm}{\hfil} }
\caption{X--ray sources detected in the central $40'\times40'$ of A194 
field overlaid on the relevant DSS plate. The radii of circles represent 
the $3\,\sigma$ confidence radius on the central position of the X--ray source.
The image has been corrected for boresight errors 
($\sim 8''$; see text).}
\label{a194poss}
\end{figure*}

These problems are simultaneously solved with a decimation process which
leads to a subset of coefficients with properties analogous to
monodimensional orthogonal wavelets.
Named $\tilde a$ the most significant scale for a detected source, we consider
only coefficients taken at the scales $\{\tilde
a/2,\tilde a,2\tilde a\}$. 
We do not extend the analysis to additional
scales since smaller scales ($< \tilde a/2$) are dominated by noise, while
larger ones ($> 2\tilde a$) are affected by source confusion.
From the WT at these three scales we then extract a subset of coefficients 
with a spacing roughly equal to the correlation length of the WT at a
particular scale and a spatial support matching the one 
of the mother at the most 
significant scale. 
In addition the denser sampling at lower scales allows any aliasing problem
to be avoided.

This new technique presents different advantages with respect to the
ones already used in X--ray analysis, i.e. 
maximum fitting (see GAL) or scale by scale
$\chi^2$ minimisation (see Rosati 1995). First the 4-dimensional fit
provides us with a fast, consistent and unique determination of all
parameters without the loss of information involved in the maximum
fitting technique or the {\it a posteriori} weighted sum of different
scale coefficients needed in the scale by scale characterisation.
Second, the measurement of the positions can be refined during the fitting 
procedure (unlike in 
the GAL technique) and the presence of nearby detections is easily
treated with a simultaneous multi-source fitting. Finally, the errors
in the estimated parameters are automatically obtained from the covariance
matrix without referring to $\chi^2$ space. This fundamental result is
due to the fact that the subset of the WT coefficients obtained after
the decimation is almost orthogonal
and Gaussian distributed.

A large set of simulations has been carried out in order to test the 
reliability of the source parameters and related errors 
(Lazzati et al. 1997).
These simulations have revealed that as long as the WT coefficients are 
sufficiently Gaussian distributed, the procedure works very well. 
With background values below $\sim 5\times 10^{-2}$ counts pixel$^{-1}$ the 
distribution in the lowest scales of WT space 
becomes increasingly Poissonian and the covariance matrix gives underestimated
errors, even if parameter determination remains reliable.

A rough distinction between point-like and extended sources can be made by 
comparing the source width (FWHM), as derived by the fitting procedure,
and the PSPC PSF at a given off-axis angle (see Figure~\ref{psfsou}). 
The solid line is a model of the behaviour of the PSPC PSF FWHM taken 
from Rosati (1995). Sources with a FWHM larger that the local PSF FWHM 
at a $>3\,\sigma$ level are classified as extended in 
Tables~\ref{a1367tab} and~\ref{a194tab}. 
It is important to stress that, since this model is valid only inside the 
PSPC support ring (vertical dashed line in Figure~\ref{psfsou}), 
classification of sources at larger angles is questionable. In 
Tables~\ref{a1367tab} and~\ref{a194tab} these classifications are
marked with a \$.

A spectral analysis has also been performed in order to better
characterise the physical properties of the X--ray sources.
Due to the low signal-to-noise of these objects only
colour--colour diagrams have been constructed rather than X--ray spectra 
(see Figure~\ref{coldiag}).
In this Figure, sources associated with cluster members (i.e.
positionally coincident with galaxies, see below) are marked with an
asterisk. Even though the colour information is not sufficient to fully
discriminate between different spectral properties, we note that the
softness--ratio of marked sources is lower than the mean value of the
sample and is consistent with the value from a source with a
bremsstrahlung spectrum with $T\lsim 2$ keV, absorbed with the
galactic column density in the direction of the target.

 \section{Individual clusters}
\label{individual}

\subsection{A1367}
\label{a1367}

\begin{table*}[!htpb]
\begin{small}
\caption{Parameters of the small-scale features detected in A1367 field.}
\label{a1367tab}
\begin{center}
\begin{tabular}{l|ccccccccc}
Name & Sign.  &      R.A.                         & DEC.       & Error box$^a$& FWHM & Count rate$^b$ & Class.$^c$ & Other work & Optical\\
     & $(\sigma)$ & (J2000) & (J2000)  & (arcsec) & (arcsec) & (c ks$^{-1}$)     &   & sources$^d$ & identifications$^e$\\ \hline
\ 1 & 32.6& 11$^{\rm h}$ 45$^{\rm m}$ 05.0$^{\rm s}$ &  19\deg 36$'$ 22$''$ &  0.8& $38\pm2$   &$256\pm8$ & P     &P8 G24 &54\\
\ 2 & 8.3& 11$^{\rm h}$ 45$^{\rm m}$ 01.0$^{\rm s}$  &  19\deg 12$'$ 49$''$ &  3.2& $160\pm5$  & $47\pm2$ &P$^\$ $&       &\\
\ 3 & 7.1& 11$^{\rm h}$ 46$^{\rm m}$ 24.0$^{\rm s}$  &  19\deg 48$'$ 28$''$ &  4.0& $114\pm7$  & $18\pm2$ &P$^\$ $&       &\\
\ 4 & 7.1& 11$^{\rm h}$ 42$^{\rm m}$ 55.9$^{\rm s}$  &  20\deg 00$'$ 38$''$ &  5.1& $201\pm10$ & $41\pm3$ &E$^\$ $&       &\\
\ 5 & 6.1& 11$^{\rm h}$ 43$^{\rm m}$ 43.7$^{\rm s}$  &  20\deg 05$'$ 54$''$ &  5.6& $150\pm10$ & $30\pm3$ &E$^\$ $&E10    &\\
\ 6 & 6.0& 11$^{\rm h}$ 45$^{\rm m}$ 40.0$^{\rm s}$  &  19\deg 53$'$ 14$''$ &  4.8& $84\pm10$  & $9\pm1$  & P     &G28    &\\
\ 7 & 5.8& 11$^{\rm h}$ 43$^{\rm m}$ 47.7$^{\rm s}$  &  19\deg 58$'$ 20$''$ &  5.6& $130\pm11$ & $33\pm4$ &E$^\$ $&G4     &$24''$ from 8\\
\ 8 & 5.3& 11$^{\rm h}$ 45$^{\rm m}$ 47.2$^{\rm s}$  &  19\deg 46$'$ 26$''$ &  5.2& $61\pm11$  & $7\pm1$  & P     &G30    &$\dag$\\
\ 9 & 4.8& 11$^{\rm h}$ 45$^{\rm m}$ 06.4$^{\rm s}$  &  19\deg 58$'$ 23$''$ &  6.4& $67\pm14$  & $5\pm1$  & P     &G23    &46\\
10 & 4.7& 11$^{\rm h}$ 44$^{\rm m}$ 35.4$^{\rm s}$  &  19\deg 51$'$ 08$''$ &  8.4& $63\pm16$ & $7\pm2$  & E     &E6 G12 &\\
11 & 4.7& 11$^{\rm h}$ 44$^{\rm m}$ 40.3$^{\rm s}$  &  19\deg 42$'$ 43$''$ &  7.2& $71\pm9$  & $17\pm3$ & E     &G14-15-17 &\\
12 & 4.6& 11$^{\rm h}$ 43$^{\rm m}$ 29.7$^{\rm s}$  &  19\deg 41$'$ 51$''$ &  6.7& $71\pm14$ & $5\pm1$  & P     &G2     &\\
13 & 4.6& 11$^{\rm h}$ 44$^{\rm m}$ 01.0$^{\rm s}$  &  19\deg 32$'$ 52$''$ &  4.2& $54\pm10$ & $6\pm1$  & P     &G7     &\\
14 & 4.6& 11$^{\rm h}$ 43$^{\rm m}$ 25.6$^{\rm s}$  &  19\deg 34$'$ 43$''$ &  8.0& $111\pm14$& $9\pm2$  &E$^\$ $&G1     &\\
15 & 4.5& 11$^{\rm h}$ 45$^{\rm m}$ 17.4$^{\rm s}$  &  19\deg 50$'$ 19$''$ & 11.0& $64\pm29$ & $3\pm2$  & P     &G26    &$45''$ from 50\\
16 & 4.5& 11$^{\rm h}$ 44$^{\rm m}$ 49.0$^{\rm s}$  &  19\deg 47$'$ 50$''$ &  5.7& $62\pm20$ & $8\pm3$  & E     &E8 G18 &35\\
17 & 4.4& 11$^{\rm h}$ 45$^{\rm m}$ 30.6$^{\rm s}$  &  19\deg 16$'$ 41$''$ &  4.9& $94\pm10$ & $6\pm2$  &P$^\$ $&       &\\
18 & 4.4& 11$^{\rm h}$ 44$^{\rm m}$ 40.5$^{\rm s}$  &  19\deg 18$'$ 32$''$ &  8.2& $104\pm18$& $7\pm3$  &P$^\$ $&       &\\
19 & 4.3& 11$^{\rm h}$ 42$^{\rm m}$ 52.6$^{\rm s}$  &  19\deg 40$'$ 18$''$ &  4.8& $80\pm10$ & $6\pm1$  &P$^\$ $&       &\\
20 & 4.2& 11$^{\rm h}$ 44$^{\rm m}$ 16.0$^{\rm s}$  &  19\deg 59$'$ 07$''$ &  8.0& $66\pm15$ & $7\pm2$  & P     &       &\\
21 & 4.2& 11$^{\rm h}$ 43$^{\rm m}$ 57.7$^{\rm s}$  &  19\deg 53$'$ 39$''$ &  5.6& $41\pm15$ & $6\pm2$  & P     &G5     &10\\
22 & 4.1& 11$^{\rm h}$ 46$^{\rm m}$ 20.5$^{\rm s}$  &  19\deg 19$'$ 02$''$ &  8.4& $124\pm14$& $6\pm1$  &P$^\$ $&       &\\
23 & 4.1& 11$^{\rm h}$ 44$^{\rm m}$ 59.7$^{\rm s}$  &  20\deg 10$'$ 55$''$ &  8.0& $113\pm14$& $7\pm1$  &P$^\$ $&       &\\
24 & 4.1& 11$^{\rm h}$ 42$^{\rm m}$ 44.2$^{\rm s}$  &  19\deg 37$'$ 49$''$ &  5.4& $86\pm10$ & $6\pm1$  &P$^\$ $&       &\\
25 & 4.0& 11$^{\rm h}$ 45$^{\rm m}$ 15.9$^{\rm s}$  &  19\deg 57$'$ 26$''$ &  5.4& $16\pm30$ & $3\pm1$  & P     &G25    &\\
26 & 4.0& 11$^{\rm h}$ 44$^{\rm m}$ 09.2$^{\rm s}$  &  19\deg 49$'$ 59$''$ & 19.6& $73\pm33$ & $4\pm3$  & P     &E2 G10 &\\
27 & 4.0& 11$^{\rm h}$ 45$^{\rm m}$ 59.6$^{\rm s}$  &  19\deg 39$'$ 52$''$ & 12.4& $86\pm26$ & $4\pm3$  &P$^\$ $&       &\\
28$^\ast$ & 15.5 & 11$^{\rm h}$ 43$^{\rm m}$ 59.3$^{\rm s}$ & 19\deg 56$'$ 56$''$ &  2.8& $125\pm5$ & $84\pm5$& - & P1 G6-8  &\\
\end{tabular}
\end{center}
\end{small}
$a$) Positional accuracy at a $1\,\sigma$ level.

$b$) Errors are computed adding quadratically the Poisson errors and the error
resulting from the fitting procedure (note that the source significance quoted 
in the second column is based only on the signal-to-noise ratio in the WT space).

$c$) P=point-like, E=extended. The presence of a $\$ $ indicates that the source is 
outside the PSPC rib, so that the classification cannot established firmly.
 
$d$) G = Grebenev et al. (1995); P = Bechtold et al. (1983) point sources; 
E = Bechtold et al. (1983) extended sources.
The identification has been carried out by combining the $3\,\sigma$ 
positional errors of the relevant catalogs.

$e$) Numbers refer to galaxies reported in Table 5 of BAL. Identifications
are at a $3\,\sigma$ level.
 
$^\ast$ This source is problematic: for GAL is double, while in our analysis
appears as a highly asymmetric feature. For this reason the characterisation
is uncertain and the source has not been included in the sample.

$^\dag$ Even if this source appears correlated with a galaxy 
(see Figure~\ref{a1367poss}), this galaxy has not been included in 
the BAL sample.

\end{table*}

A1367 is a nearby cluster ($z=0.022$) well known for its very irregular 
morphology. Already twelve years ago {\it Einstein} images were used to
search for X--ray emission from cluster galaxies (BAL). 
The more recent analysis of GAL has confirmed some of the {\it
Einstein} sources and discovered new ones.
Our analysis of the 18224~s PSPC observation reveals the presence of 28 
sources within three times the major and minor axis core-radii of the King 
profile, with a  significance of $4\,\sigma$.  
It is relevant for our study to compare these surface densities with the those
typical of the field. Using the Hasinger et al. (1993)
$\log N - \log S$ relation, we would expect only three to nine sources
within the same solid angle and down to the flux limit of
$\sim 5\times10^{-14}$~erg~cm$^{-2}$~s$^{-1}$.
In Table~\ref{a1367tab} we report the sources detected in the field, along with
statistical significance, position, size, count rate (and relative errors) as 
well as extent classification and coincidences with the detections of GAL 
and BAL. Only a small number of BAL's sources (6/21) has been 
re-discovered despite the fact that the {\it Einstein} field of view is 
completely contained in the ROSAT PSPC image. 
GAL, who used the same data of this work, found 13 out of 21 BAL sources.
We suspect that a sizeable number of the small-scale features detected by BAL
are a result of their technique which has difficulties in assessing the
significance of the detections.
A larger number of sources detected in this work 
coincide with those of GAL (16 
out of 28, even if they adopted a significance level of $3\,\sigma$
and perform the search in a smaller area). 
However, many of the sources classified as extended by GAL, have been 
found to be point-like in the present work. This difference is due 
to the extended emission of the cluster which has not been accounted for by 
GAL (see subsection 3.2) and could explain the lack of identifications of 
extended sources with galaxies.

\begin{table*}[!htpb]
\caption{Parameters of the small-scale features detected in A194 field.}
\label{a194tab}
\centerline{
\begin{tabular}{l|cccccccc}
Name&Sign.& R.A.         & Dec.         & Error Box$^a$& FWHM & Count rate$^b$ & Class$^c$ & Optical\\
   &$(\sigma)$& (J2000)  &  (J2000)     &(arcsec)& (arcsec)& (c ks$^{-1}$) &        & identifications$^d$\\ \hline
\ 1  & 14.1 & 1$^{\rm h}$ 27$^{\rm m}$ 24.9$^{\rm s}$ & --1\deg 44$'$ 07$''$ & 2.7&$163\pm5$  &$112\pm4$ & P$^\$ $&\\
\ 2  & 12.9 & 1$^{\rm h}$ 26$^{\rm m}$ 42.0$^{\rm s}$ & --1\deg 14$'$ 15$''$ & 1.0& $58\pm2$  &$115\pm4$ & P      &\\
\ 3  & 12.3 & 1$^{\rm h}$ 26$^{\rm m}$ 00.0$^{\rm s}$ & --1\deg 20$'$ 48$''$ & 1.2& $43\pm3$  & $94\pm4$ & P      &NGC547\\
\ 4  & 10.0 & 1$^{\rm h}$ 25$^{\rm m}$ 35.7$^{\rm s}$ & --1\deg 25$'$ 54$''$ & 1.8& $64\pm4$  & $62\pm4$ & E      &\\
\ 5  &  9.4 & 1$^{\rm h}$ 27$^{\rm m}$ 09.1$^{\rm s}$ & --1\deg 52$'$ 35$''$ & 4.8&$202\pm9$  & $80\pm5$ & P$^\$ $&\\
\ 6  &  7.8 & 1$^{\rm h}$ 27$^{\rm m}$ 00.8$^{\rm s}$ & --1\deg 30$'$ 09$''$ & 2.4& $86\pm5$  & $36\pm2$ & P$^\$ $&\\
\ 7  &  7.2 & 1$^{\rm h}$ 26$^{\rm m}$ 11.3$^{\rm s}$ & --1\deg 12$'$ 05$''$ & 1.7& $40\pm4$  & $28\pm2$ & P      &\\
\ 8  &  7.1 & 1$^{\rm h}$ 26$^{\rm m}$ 22.3$^{\rm s}$ & --1\deg 38$'$ 13$''$ & 3.2& $82\pm6$  & $23\pm2$ & P      &\\
\ 9  &  7.0 & 1$^{\rm h}$ 25$^{\rm m}$ 15.4$^{\rm s}$ & --1\deg 05$'$ 13$''$ & 3.2& $91\pm6$  & $29\pm2$ & P$^\$ $&\\
10 &  6.7 & 1$^{\rm h}$ 23$^{\rm m}$ 52.7$^{\rm s}$ & --1\deg 17$'$ 50$''$ & 6.2&$133\pm12$ & $22\pm3$ & P$^\$ $&\\
11 &  6.7 & 1$^{\rm h}$ 25$^{\rm m}$ 30.0$^{\rm s}$ & --1\deg 37$'$ 28$''$ & 3.6& $87\pm7$  & $20\pm2$ & E      &\\
12 &  6.6 & 1$^{\rm h}$ 25$^{\rm m}$ 44.1$^{\rm s}$ & --1\deg 22$'$ 49$''$ & 2.4& $49\pm6$  & $31\pm3$ & P      &NGC541\\
13 &  6.5 & 1$^{\rm h}$ 25$^{\rm m}$ 41.7$^{\rm s}$ & --1\deg 44$'$ 25$''$ & 6.4&$152\pm10$ & $32\pm3$ & E$^\$ $&Cluster \\
14 &  6.3 & 1$^{\rm h}$ 25$^{\rm m}$ 09.2$^{\rm s}$ & --1\deg 35$'$ 33$''$ & 3.6& $77\pm7$  & $17\pm2$ & P      &\\
15 &  6.0 & 1$^{\rm h}$ 24$^{\rm m}$ 49.8$^{\rm s}$ & --1\deg 23$'$ 40$''$ & 3.6& $71\pm7$  & $17\pm2$ & P      &\\
16 &  5.9 & 1$^{\rm h}$ 24$^{\rm m}$ 57.4$^{\rm s}$ & --1\deg 17$'$ 37$''$ & 2.4& $44\pm6$  & $13\pm1$ & P      &\\
17 &  5.8 & 1$^{\rm h}$ 24$^{\rm m}$ 30.0$^{\rm s}$ & --1\deg 26$'$ 47$''$ & 3.2& $81\pm6$  & $25\pm2$ & P$^\$ $&\\
18 &  5.4 & 1$^{\rm h}$ 27$^{\rm m}$ 45.7$^{\rm s}$ & --1\deg 52$'$ 59$''$ & 9.6&$158\pm20$ & $23\pm5$ & P$^\$ $&NGC564\\
19 &  5.1 & 1$^{\rm h}$ 25$^{\rm m}$ 55.4$^{\rm s}$ & --1\deg 11$'$ 23$''$ & 3.2& $37\pm8$  & $12\pm2$ & P      &\\
20 &  4.9 & 1$^{\rm h}$ 24$^{\rm m}$ 17.5$^{\rm s}$ & --1\deg 14$'$ 46$''$ &10.4&$147\pm23$ & $15\pm4$ & E$^\$ $&Cluster\\
21 &  4.7 & 1$^{\rm h}$ 27$^{\rm m}$ 27.6$^{\rm s}$ & --1\deg 03$'$ 57$''$ &10.4&$135\pm15$ & $13\pm2$ & P$^\$ $&\\
22 &  4.6 & 1$^{\rm h}$ 24$^{\rm m}$ 20.0$^{\rm s}$ & --1\deg 02$'$ 27$''$ &11.6&$191\pm21$ & $23\pm4$ & E$^\$ $&\\
23 &  4.5 & 1$^{\rm h}$ 26$^{\rm m}$ 30.2$^{\rm s}$ & --1\deg 53$'$ 44$''$ &11.6&$158\pm24$ & $15\pm4$ & P$^\$ $&\\
24 &  4.4 & 1$^{\rm h}$ 24$^{\rm m}$ 44.3$^{\rm s}$ & --0\deg 52$'$ 38$''$ & 8.0&$94\pm15$  & $8\pm2$  & P$^\$ $&\\
25 &  4.2 & 1$^{\rm h}$ 25$^{\rm m}$ 08.2$^{\rm s}$ & --0\deg 52$'$ 59$''$ & 4.4& $54\pm9$  & $8\pm1$  & P$^\$ $&\\
26 &  4.2 & 1$^{\rm h}$ 25$^{\rm m}$ 16.0$^{\rm s}$ & --1\deg 31$'$ 31$''$ & 9.6&$126\pm16$ & $14\pm3$ & E      &NGC538\\
\end{tabular}}
$a$) Positional accuracy at a $1\,\sigma$ level.

$b$) P=point-like, E=extended. The presence of a $\$ $ indicates that the source is 
outside the PSPC rib, so that the classification cannot established firmly.

$c$) Errors are computed adding quadratically the Poisson errors and the error
resulting from the fitting procedure (note that the source significancy quoted
in the second column is based only on the signal-to-noise ratio in the WT space).

$d$) The identification has been carried out by using a $3\,\sigma$ 
positional error in the NGC galaxy catalog.
 
\end{table*}

\begin{figure}[!ht]
\psfig{figure=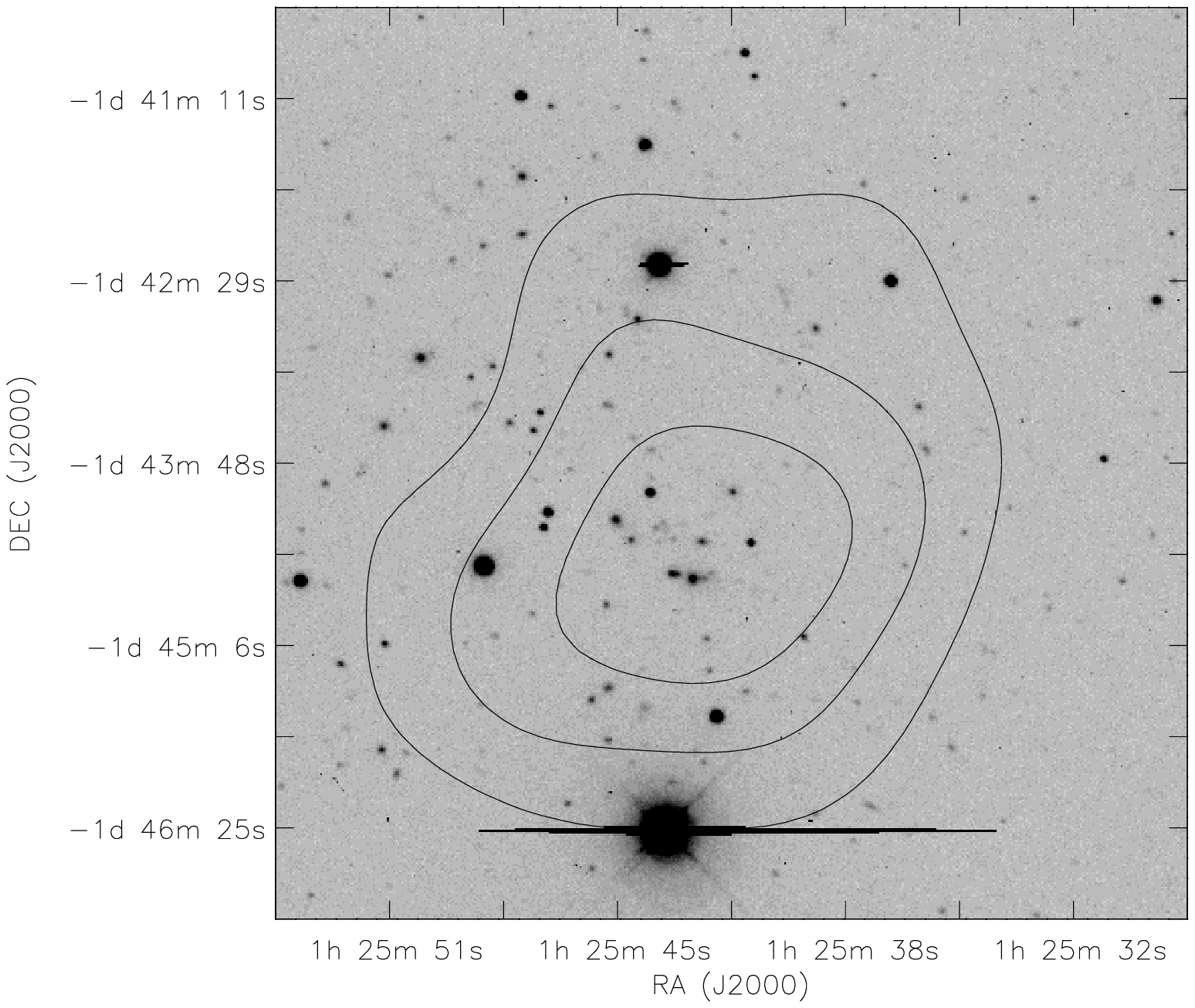,width=9cm}
\caption{X--ray contours in the ROSAT hard band of source 13 in A194 field 
overlaid on a
CCD R band image (15 min exposure at the CTIO 1.5~m).}
\label{over13}
\end{figure}

\begin{figure}[!hb]
\psfig{figure=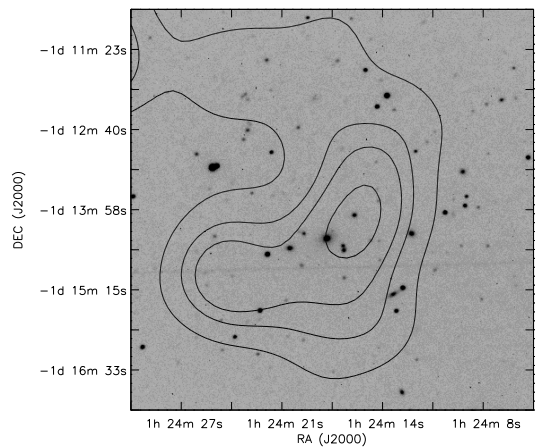,width=9cm}
\caption{X--ray contours in the ROSAT hard band of source 20 in A194 field 
overlaid on a
CCD I band image (15 min exposure at the CTIO 1.5~m).}
\label{over20}
\end{figure}

For this purpose in Figure~\ref{a1367poss} the positions of the 
detected sources are overlaid on a $40'\times 40'$ DSS
plate of the A1367 center:
radii of circles are equal to the $3\,\sigma$ error boxes.
No secure identifications with stars were found for the boresight 
correction so that we only used the bright elliptical galaxy  
displaced by $\sim 14''$ from the brightest X--ray source (1) (see 
Figure~\ref{a1367poss}).
This value has been adopted for the boresight correction and provides
a good match also for the other sources (e.g. 8, 11, 13).
Out of the 28 detected X--ray sources, 3 have galaxies members of the 
cluster within their error boxes (and possibly other 2, see 
Table~\ref{a1367tab}) and 1 is a quasar (source 28). 
The identification of these galaxies with cluster members is further
supported by the fact that their fluxes (calculated using a 2~keV
bremsstrahlung spectrum and a $5\times10^{20}$~cm$^{-2}$ column
density) would convert into luminosities ranging from 0.6 to
$6.6\times 10^{41}$ erg s$^{-1}$ at the cluster redshift.
GAL presented several more identifications, but some appear to be inconsistent 
with the better positional accuracy obtained with  
our technique (see Table~\ref{a1367tab}). 
We also find several extended sources (4, 5, 7, 10, 11, 14, 16 and 28, see 
Table~\ref{a1367tab}). The extension of source 11 may 
be still contaminated by the cluster emission, being at its center, 
while source 28 is problematic (see note in Table~\ref{a1367tab}).
The classification of sources outside the PSPC rib however is less reliable.

\subsection{A194}
\label{A194}

Abell 194 is a nearby ($z=0.018$) poor ($R=0$) cluster with little
extended emission. The 16317 s ROSAT observation had not been previously
analysed. 
The results are reported in Table~\ref{a194tab}.
In Figure~\ref{a194poss} the X--ray sources are overlaid on a $40'\times 
40'$ DSS plate of the A194 center. In this case a boresight correction 
of $8''$ was applied, taking as a reference three Guide Star Catalog objects 
(source 2, 12 and 17 in Table~\ref{a194tab}).
We detected 26 sources 4 of which are known galaxies of the cluster.
In this case the expected number of background object is even lower
(given the higher flux limit) than in the case of A1367.

The comparison of the source extension (FWHM) with the
PSPC PSF allows us to detected 6 extended sources, namely 4, 11, 13, 20,
22 and 26 (see Table~\ref{a194tab} and Figure~\ref{psfsou}). Two of these
sources (12 and 26) are likely associated with galaxies members of
the cluster. Follow-up CCD observations obtained at the CTIO 1.5~m telescope
reveal a significant overdensity of galaxies around sources 13 and 20
(Figures \ref{over13} and \ref{over20}). Thus, we believe that these sources
are two newly discovered intermediate redshift galaxy clusters.

\section{Summary}
\label{concl}

We have presented a new technique for a multi-scale analysis of
astronomical X--ray images particularly suited for the detection and
the characterisation of small and intermediate scale features embedded
in a strongly varying background, such as the emission from clusters of
galaxies.
We have showed that even a rough subtraction of the extended component
is needed in order to avoid positional biases in the detection
thresholds.  Our technique makes use of a 4-dimensional fit in the
small and intermediate scale wavelet space which is first ``decimated''
to avoid redundancy and strong correlations.
This approach leads to a better characterisation of the detected
sources when compared with a 2-dimensional fit over the entire range of
scales (e.g. GAL).

The application of this technique to two nearby cluster of galaxies 
has revealed a sizeable number of small-scale features which are thus 
believed to be a common property of nearby clusters.
We detected 28 sources in the central part of A1367 (up to about $30'$
from its center) and 26 in A194 ($28'$ from its center).
Since this surface density of sources significantly exceeds the value
expected from the number counts in the field, we conclude that most of
these sources are indeed physically bound to the cluster.

We provide a catalog of sources indicating (when possible) their likely
optical identifications and those classified as extended. Two of the
extended sources in A194 field are found to be associated with
serendipitous background galaxy clusters expected to lie
at intermediate redshifts.

\begin{acknowledgements}
This work has utilised images from the Digitized Sky Survey.
The Digitized Sky Surveys were produced at the Space
Telescope Science Institute under U.S. Government grant NAG W-2166.
We thank the anonymous referee for
providing useful comments and suggesting important improvements.
\end{acknowledgements}

\end{document}